\documentclass[10pt]{iopart}

\usepackage{iopams}

\expandafter\let\csname equation*\endcsname\relax
\expandafter\let\csname endequation*\endcsname\relax
\usepackage{amsmath}
\usepackage{graphicx}

\usepackage{xcolor}
\usepackage{soul}

\begin{document}

\title[]{Quasi-harmonic thermoelasticity of palladium, platinum, copper, and gold
from first principles}

\author{Cristiano Malica$^1$ and Andrea Dal Corso$^{1,2}$}
\address{$^1$International School for Advanced Studies (SISSA), \\
\ \ Via Bonomea 265, 34136 Trieste (Italy).}

\address{$^2$CNR-IOM, \\
\ \ Via Bonomea 265, 34136 Trieste (Italy).}

\ead{cmalica@sissa.it}

\begin{abstract}
We calculate the temperature-dependent elastic constants 
of palladium, platinum, copper and gold within the quasi-harmonic approximation
using a first-principles approach and evaluating numerically the second derivatives 
of the Helmholtz free-energy with respect to strain at the minimum of the free-energy itself.
We find an overall good agreement with the experimental data although
the anomalies of palladium and platinum reported at room temperature are
not reproduced.
The contribution of electronic excitations is also investigated: we find that
it is non-negligible for the $C_{44}$ elastic constants
of palladium and platinum while it is irrelevant in the other cases. Its effect
is not sufficient to explain the details of the anomalies found by experiments,
not even when, in the case of platinum, we take into account the electron-phonon
interaction.
Lastly, the effect of the exchange and correlation functional is addressed
and it is found that it is important at $T=0$ K, while all functionals
give similar temperature dependencies.
\end{abstract}

%
%
%
%
\ioptwocol

\section{Introduction}
The vibrational and thermodynamic properties of transition and noble metals
and of their alloys are key ingredients to understand many industrial
and technological processes.~\cite{Saunders_1998, Lukas_2007, Curtarolo_2012}
Moreover, the comprehension of these properties for elemental metals is
a fundamental prerequisite for the rational design of materials.
In addition to the extensive experimental investigations~\cite{Touloukian_1975},
the prediction of thermodynamic properties via density functional
theory (DFT) ~\cite{Hohenberg_1964, Kohn_1965} has been among the primary goals
of first-principles studies for many years (see for instance
Refs.~\cite{Grabowski_2007, Dal_Corso_2013} and references therein).

In this context, the \texttt{thermo\_pw}~\cite{tpw} code has
been designed to compute efficiently, among other things, the thermodynamic
properties of solids. Previous applications include the phonon dispersions, the
thermal expansions, and the heat capacities of the hexagonal close packed (hcp) 
metals rhenium, technetium,~\cite{pal2} ruthenium, and osmium~\cite{pal} and the anharmonic
contributions to the mean square atomic displacements within the quasi-harmonic
approximation (QHA) of silicon, ruthenium, magnesium, and cadmium~\cite{bf}.

Elastic constants (ECs) are other crucial quantities for
crystal thermodynamics: they determine the crystal stability,
thermal stresses, and sound velocities. For several decades
DFT has provided ECs of solids, often within $10 \%$
from the experiment. However calculations have been usually limited
to $T=0$ K since the introduction of quantum and temperature effects
requires a significant computational effort. As a result,
there are now numerous theoretical ECs data
at $T=0$ K~\cite{Wang_2009, Golesorkhtabar_2013, Jamal_2014},
but much less at finite temperatures.
Some calculations do exist, but many of them are limited to the quasi-static
approximation (QSA) where the temperature dependent elastic constants
(TDECs) are calculated as second derivatives of the total energy with
respect to strain (as at $T=0$ K) at the minimum of the Helmholtz
free-energy so accounting only for the effect of thermal expansion on the ECs.
Only in a few papers the complete QHA method was applied (see, for
example, Refs. ~\cite{karki, dragoni}).
Other calculations make use of the $ab$-$initio$ molecular dynamics
method that, however, focus mainly on the high temperature behavior where quantum
effects can be neglected~\cite{Steneteg_2013}.

The \texttt{thermo\_pw} code can compute the ECs at $T=0$ K
(see for instance \cite{Dal_Corso_2016} and references therein)
and, recently, the implementation was extended to the calculation of the quasi-harmonic
TDECs as second derivatives of the Helmholtz free-energy with respect
to strain~\cite{Malica_2020}. So far, applications include the TDECs
of silicon, aluminum, silver,~\cite{Malica_2020} and boron arsenide~\cite{Malica_2020_b}.

In this work, we apply this approach to other four paradigmatic 
face-centered cubic (fcc) transition and noble metals: palladium, platinum, copper, and gold.
The first two elements, palladium and platinum, present anomalies in the
experimental TDECs that are not well understood, while the latter
two are more regular.
Already after the early measurements on palladium and platinum some models were put
forward to explain the anomalous temperature dependence of their elastic
constants~\cite{Rayne_1960, Weinmann_1974, Yoshihara_1987, 
Collard_1992} and recently a computational DFT study has supported
this interpretation~\cite{Keuter_2019}.
These studies pointed out that the TDECs of palladium and platinum are anomalous due to
the partially filled $d$ bands whose electrons contribute
substantially to the free-energy. This theoretical study focused on the
electronic contribution to the TDECs while the phonon contribution was
accounted for within the QSA. Another theoretical study within the QSA,
instead, finds a conventional temperature dependence for all the ECs of palladium, even
considering the electronic thermal excitations~\cite{Liu_2011}.

Here we extend the previous QHA implementation in order to include
the effects of the electronic excitations on the TDECs and
compare the electronic and vibrational contributions.
We find that the electronic contribution, although smaller than the vibrational
one, is relevant for the temperature dependence of the $C_{44}$ ECs of
palladium and platinum, where it improves the comparison with experiment,
but it is not sufficient to reproduce in detail the anomalies.
On the contrary, in copper and gold, whose $d$ shells are completely filled,
the electronic contribution to the TDECs is negligible.

Focusing on platinum we investigate also another possible source of anomalous
behavior. Actually electronic excitations might change the phonon frequencies
especially at high temperatures, so we calculate the effect of
this change on the TDECs. We find no substantial contribution, not even at
the highest studied temperature of $T=1000$ K.

Finally, since so far the effect of the exchange and correlation functionals
on the TDECs has not been addressed in detail, we do also a systematic
comparison between the local density approximation (LDA) and
the generalized gradient approximation (GGA), taking the available experimental
data as a reference.
We find that all functionals give similar softenings of the ECs with temperature,
hence the functional that matches better the experimental ECs at $T=0$ K
turns out to be also the one that matches better the TDECs.

\section{Theory}
The QHA approach to the TDECs is detailed in the
recent works~\cite{Malica_2020, Malica_2020_b}: in this Section we limit ourselves to a summary
of the most important formulas in order to make the paper self-contained and
discuss how we have introduced the electronic excitations in the calculation.

The isothermal elastic constants are obtained from the derivatives of the Helmholtz
free-energy $F$ with respect to strain $\epsilon$:
\begin{equation}\label{dUT}
\tilde C_{ijkl}^T = \frac{1}{\Omega} \left(\frac{\partial^2 F}{\partial \epsilon_{ij} 
\partial \epsilon_{kl}} \right)_{\epsilon=0},
\end{equation}
where $\Omega$ is the unit cell volume. 
 
From the previous Equation the free-energy $F$ contains a term quadratic in
the strains:
\begin{equation} \label{u_en}
F = \frac{\Omega}{2} \sum_{ijkl} \epsilon_{ij} \tilde C_{ijkl}^T \epsilon_{kl}.
\end{equation}
In cubic solids there are three independent ECs, $C_{1111}$, $C_{1122}$
and $C_{2323}$ which, for symmetry reasons~\cite{nyebook} are usually written in Voigt's
notation as $C_{11}$, $C_{12}$ and $C_{44}$, respectively. 
We compute them using the deformations:
\begin{equation} \label{strains}
\begin{aligned}
\epsilon_A & =\left( \begin{array}{ccc}
\epsilon_{1} & 0 & 0 
\\
0 & \epsilon_{1} &  0 
\\
0 & 0 & \epsilon_{1}
\end{array}
\right),\ \ 
\epsilon_{E}=\left( \begin{array}{ccc}
0 & 0 & 0 
\\
0 & 0 &  0
\\
0 & 0 & \epsilon_{3}
\end{array}
\right), \\
\epsilon_{F} & =\left( \begin{array}{ccc}
0 & \epsilon_{4} & \epsilon_{4} 
\\
\epsilon_{4} & 0 & \epsilon_{4}
\\
\epsilon_{4} & \epsilon_{4} & 0
\end{array}
\right).
\end{aligned}
\end{equation}
The strain $\epsilon_A$ does not change the shape of the cubic cell, 
while $\epsilon_E$ transforms it into a tetragonal cell and $\epsilon_F$
into a rhombohedral cell. None of them conserves the volume of the 
cell.

As we are interested in the ECs obtained from the stress-strain
relationship ($C_{ijkl}^T$ without $tilde$), when the system is under a pressure $p$,
we correct the $\tilde C_{ijkl}^T$~\cite{Barron_1965}:
\begin{equation}\label{dUp}
C^T_{ijkl} = \tilde C^T_{ijkl} + \frac{1}{2} p \left(2 \delta_{ij} \delta_{kl} - \delta_{il} \delta_{jk} 
- \delta_{ik} \delta_{jl} \right).
\end{equation}
The Helmholtz free-energy of Eq.~\ref{u_en} is obtained as the sum of the DFT total
energy $U$, the vibrational and the electronic free energies $F=U+F_{vib}+F_{el}$.
The vibrational free-energy is given by:
\begin{eqnarray}\label{fvib}
F_{vib}(\mathbf \epsilon, T) =&&\frac{1}{2N} \sum_{\mathbf q \eta} \hbar \omega_{\eta} 
\left(\mathbf q, \mathbf \epsilon \right) + \nonumber\\
&&+ \frac{k_B T}{N} \sum_{\mathbf q \eta} \ln \left[1 - \exp \left(-\frac{\hbar \omega_{\eta}
(\mathbf q, \mathbf \epsilon)}{k_B T}\right) \right],
\end{eqnarray}
where $N$ is the number of cells in the crystal,
$\omega_{\eta}\left(\mathbf q, \epsilon \right)$ is the phonon angular frequency
of the mode $\eta$ with wave-vector $\mathbf q$ computed in the system with a
strain $\epsilon$. The $\omega_{\eta}\left(\mathbf q, \epsilon \right)$ are
computed by density functional perturbation theory (DFPT)~\cite{rmp} on
a coarse ${\bf q}$-point mesh and Fourier interpolated on a thicker mesh
to perform the Brillouin zone summation.

In order to account for the thermal electronic excitations~\cite{Zhang_2017}, one could
use the Mermin's finite temperature extension to DFT~\cite{Mermin_1965}.
In this method, one minimizes the
free-energy $U-TS_{el}$ where both $U$ and the electronic entropy $S_{el}$
are calculated using the Fermi-Dirac occupations at a given fixed
temperature $T$.
The method is accurate but needs a self-consistent calculation for
each temperature.
In order to evaluate the electronic free-energy for many temperatures
we compute the total energy $U$ with a Methfessel-Paxton
smearing~\cite{mp} and a broadening sufficiently small to get
the zero broadening value of $U$ and then use the $T=0 $ K DFT energy bands
to calculate the contribution of electronic thermal excitations
to the free-energy: $F_{el}$. We take $F_{el}=U_{el}-TS_{el}$
where $U_{el}$ is the electronic excitations energy:
\begin{eqnarray} \label{ele_int}
U_{el}=\int_{-\infty}^{+\infty} E N(E) f dE - \int_{-\infty}^{E_{F}} E N(E) dE,
\end{eqnarray}
while $S_{el}$ is the electronic excitations entropy:
\begin{eqnarray} \label{ele_entr}
S_{el}=-k_B \int_{-\infty}^{+\infty} \left[ f \ln f + (1-f) 
\ln(1-f) \right]N(E) dE.
\end{eqnarray}
In Eqs. ~\ref{ele_int} and ~\ref{ele_entr} $f(E,\mu,T)$ are the Fermi-Dirac
occupations (depending on $E$, the temperature $T$, and
the chemical potential $\mu$), $N(E)$ is the electronic density of states and
$E_F$ is the Fermi energy. 
The $\mu$ is calculated at each temperature imposing that the number of electrons per unit 
cell $N_{el}$ is determined by $N_{el} = \int_{- \infty}^{\infty} N(E) f(E,T,\mu) dE$.

The QHA calculation of the ECs, Eq.~\ref{dUT}, is performed on a grid of reference
geometries by varying the lattice constant $a_0$.
Phonon dispersions and electronic bands are computed on the same grid in order to evaluate
the total Helmholtz free-energy as a function of the volume.
Then, at each temperature $T$, the ECs are interpolated and evaluated
at the $a(T)$ which minimizes the free-energy.
The calculation requires phonon dispersions and electronic band structures in all the
strained configurations for all the reference geometries.

In order to compare our results with ultrasonic experimental data we compute
also the adiabatic ECs given by:
\begin{equation} \label{adiab}
C_{ijkl}^S = C_{ijkl}^T + \frac{T \Omega b_{ij} b_{kl}}{C_{V}},
\end{equation}
where $b_{ij}$ are the thermal stresses:
\begin{equation}
b_{ij} = - \sum_{kl} C_{ijkl}^T \alpha_{kl},
\end{equation}
and $\alpha_{kl}$ is the thermal expansion tensor.
$C_V$ is the isochoric heat capacity defined, for example,
in Ref.~\cite{Malica_2020_b} and $\Omega$ is the unit
cell volume at the temperature $T$.

The DFPT phonon calculation can be also performed at finite temperature $T_{FD}$
by using Fermi-Dirac occupations~\cite{DeGironcoli_1995} with a broadening
corresponding to the given temperature. Hence, the angular
frequency $\omega_{\eta}\left(\mathbf q, \epsilon \right)$ of Eq.~\ref{fvib} are
replaced with the temperature dependent $\omega_{\eta}\left(\mathbf q, \epsilon, T_{FD} \right)$,
giving a correction to the Helmholtz free-energy. In this case the
electronic excitations contribution is already included in the DFT total
energy $U$, via the Mermin's functional, and it is not necessary
to add it separately.
We use this approach in platinum at enough large temperatures
in order to clearly identify the effect, in particular at the temperatures $T_{FD}=800$ 
K and $T_{FD}=1000$ K and compare with the former method.

\begin{table*} \centering
\caption{ECs at $T=0$ K computed with LDA, PBEsol and PBE exchange and correlation functionals
compared with previous theoretical works and experimental data. The equilibrium lattice constants 
($a_0$) are in \AA\ while the ECs and the bulk moduli $B$ are in kbar.
The bulk modulus is $B=\frac{1}{3}(C_{11}+2C_{12})$.
}
\begin{tabular}{lccccc}
\hline
& \multicolumn{1}{c}{$a_0$} \ \ & \multicolumn{1}{c}{$C_{11}$} \ \ & \multicolumn{1}{c}{$C_{12}$} \ \ & \multicolumn{1}{c}{$C_{44}$} \ \ & \multicolumn{1}{c}{$B$} \\
\hline
\hline
Palladium \\
\hline
LDA$$ & 3.85 & 2696 & 2071 & 788 & 2279 \\
PBEsol$$ & 3.88 & 2445 & 1861 & 740 & 2056 \\
PBE & 3.95 & 2010 & 1532 & 606 & 1690 \\
LDA$^a$ & 3.90 & 2743 & 1463 & 716 & 1890 \\
GGA$^b$ & 3.94 & 2548 & 1358 & 587 & 1755 \\
Expt.$^c$ & 3.8896(2) & 2341(27) & 1761(27) & 712(3) & 1954 \\
\hline
\hline
Platinum \\
\hline
LDA$$ & 3.90 & 3800 & 2759 & 802 & 3106 \\
PBEsol$$ & 3.92 & 3553 & 2581 & 764 & 2905 \\
PBE & 3.98 & 3039 & 2234 & 615 & 2502 \\
LDA$^d$ & 3.91 & 3645 & 2665 & 736 & 2992 \\
PBEsol$^e$ & 3.91 & 3595 & 2456 & 858 & 2836 \\
GGA$^d$ & 3.99 & 3063 & 2133 & 730 & 2443 \\
Expt.$^f$ & 3.92268(4) & 3487(130) & 2458(130) & 734(20) & 2801(9) \\
\hline
\hline
Copper \\
\hline
LDA$$ & 3.52 & 2349 & 1666 & 998 & 1894 \\
PBEsol$$ & 3.56 & 2100 & 1480 & 938 & 1687 \\
PBE$$ & 3.63 & 1775 & 1238 & 793 & 1417 \\
LDA$^g$ & 3.64 & 1678 & 1135 & 745 & 1316 \\
GGA$^h$ & 3.63 & 1745 & 1253 & 752 & 1417 \\
GGA$^i$ & 3.63 & 1800 & 1200 & 840 & 1400 \\ 
Expt.$^l$ & 3.596 & 1762.0 & 1249.4 & 817.7 & 1420.3 \\
\hline
\hline
Gold \\
\hline
LDA$$ & 4.05 & 2120 & 1873 & 373 & 1955 \\
PBEsol$$ & 4.10 & 1926 & 1651 & 366 & 2614 \\
PBE & 4.16 & 1544 & 1327 & 268 & 1389 \\
GGA$^g$ & 4.07 & 2021 & 1742 & 379 & 1835 \\
PBE$^m$ & 4.19 & 1478 & 1435 & 387 & 1449 \\
Expt.$^n$ & 4.062 & 2016.3 & 1696.7 & 454.4 & 1803.2 \\
\hline
\end{tabular}
\label{table2}
\\ $^a$ Ref.~\cite{Liu_2011}, $^b$ Ref.~\cite{Gao_2018}
\\ $^c$ Ref.~\cite{Rayne_1960} (room temperature $a_0$, $T=0$K extrapolation of ECs),
\\ $^d$ Ref.~\cite{Isaev_2008}, $^e$ Ref.~\cite{Jamal_2014} (all electrons),
\\ $^f$ Reference~\cite{Kamada_2019} (room temperature results, most recent work) 
\\ $^g$ Reference~\cite{Wang_2009}, $^h$ Reference~\cite{Liu_2012}, $^i$ Reference~\cite{Hafner_2009} 
\\ $^l$ Ref.~\cite{Haas_2009} for $a_0$ and Ref.~\cite{Overton_1955} for the ECs ($T=0$K extrapolation),
\\ $^m$ Ref.~\cite{Kong_2018}, $^n$ Ref.~\cite{Haas_2009} for $a_0$ and Ref.~\cite{Neighbours_1958} for the ECs ($T=0$K extrapolation)
\end{table*}

\begin{figure}
\centering
\includegraphics[width=\linewidth]{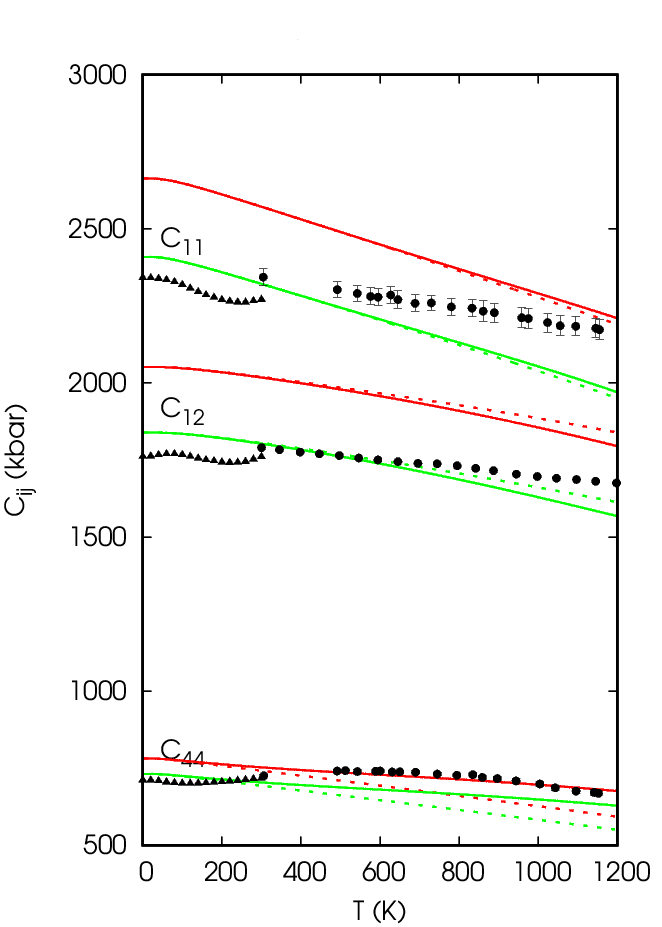}
\caption{Quasi-harmonic adiabatic elastic constants of palladium. LDA (red curves) is compared
         with PBEsol (green curves). The results obtained with the total
free-energy (continuous lines) are compared with those in which the contribution
of electronic excitations is neglected (dashed lines). Experimental data are taken from Rayne~\cite{Rayne_1960} (black triangles)
         and~\cite{Yoshihara_1987} (black circles). }
\label{fig:pd}
\end{figure}

\section{Computational parameters}
The calculations presented in this work were carried out using DFT as implemented
in the Quantum ESPRESSO package ~\cite{qe1} ~\cite{qe2}.
For the calculations of ECs at $T=0$ K and for all materials, the exchange and correlation
functional was approximated by the local density approximation (LDA)~\cite{lda}, the generalized gradient
approximation (GGA) of Perdew, Burke and Ernzerhof (PBE)~\cite{pbe} and its modification for densely
packed solids (GGA-PBEsol)~\cite{pbesol}. For the TDECs we use LDA and GGA-PBEsol for all metals and for copper 
we also use the GGA-PBE functional (which, in this case, gives the best agreement with the experimental T=0 K ECs).
We employ the projector augmented wave (PAW) method and a plane waves basis set with
pseudopotentials~\cite{paw} from \texttt{pslibrary}~\cite{psl}.
The pseudopotentials are reported in the note \footnote{For palladium we used \texttt{Pd.pz-n-}\texttt{kjpaw\_psl.1.0.0.UPF} and 
\texttt{Pd.pbesol-n-}\texttt{kjpaw\_psl.1.0.0.UPF}. 
For platinum \texttt{Pt.pz-n-}\texttt{kjpaw\_psl.1.0.0.UPF} and
\texttt{Pt.pbesol-n-}\texttt{kjpaw\_psl.1.0.0.UPF}.
For copper \texttt{Cu.pz-dn-}\texttt{kjpaw\_psl.1.0.0.UPF}, 
\texttt{Cu.pbe-dn-}\texttt{kjpaw\_psl.1.0.0.UPF} and \texttt{Cu.pbesol-dn-}\texttt{kjpaw\_psl.1.0.0.UPF}.
For gold we used \texttt{Au.pz-dn-}\texttt{kjpaw\_psl.0.3.0.UPF} and \texttt{Au.pbesol-dn-}\texttt{kjpaw\_psl.0.3.0.UPF}.}.
The cutoff for the wave functions (charge density) was $60$ Ry ($400$ Ry) for palladium, $45$ Ry ($300$ Ry) for platinum,
$60$ Ry ($1200$ Ry) for copper and $60$ Ry ($400$ Ry) for gold.
The presence of the Fermi surface has been dealt with by a smearing approach ~\cite{mp} with a smearing parameter 
$\sigma = 0.005$ Ry for palladium and platinum and $\sigma = 0.02$ Ry for copper and gold.
The \textbf{k}-point mesh was $40 \times 40 \times 40$ (except for PBEsol gold, for which a mesh $48 \times 48 \times 48$ has been used).
Density functional perturbation theory (DFPT)~\cite{rmp} ~\cite{dfptPAW} was used to calculate the dynamical matrices on a 
$8 \times 8 \times 8$ \textbf{q}-point mesh for palladium and platinum (corresponding to 29 special \textbf{q}-points for configurations 
strained with $\epsilon_{A}$, 59 \textbf{q}-points for $\epsilon_{E}$ and 65 for $\epsilon_{F}$, see Ref.~\cite{Malica_2020} for the
definitions of the strains) and $4 \times 4 \times 4$ \textbf{q}-point mesh for copper and gold (corresponding to 8 \textbf{q}-points for configurations 
strained with $\epsilon_{A}$ and 13 \textbf{q}-points for those strained with $\epsilon_{E}$ and $\epsilon_{F}$).
For palladium and platinum a thicker \textbf{q}-points mesh was necessary due to the presence of Kohn anomalies~\cite{Dal_Corso_2013}. 
For all materials the dynamical matrices have been Fourier interpolated on a $200 \times 200 \times 200$ 
\textbf{q}-point mesh to evaluate the vibrational free-energy.
The calculation of the TDECs is done by \texttt{thermo\_pw} as described in Ref.~\cite{Malica_2020}. 
The grid of reference geometries was centered at the minimum of the total energy as
reported in Table 1 except for the LDA study of platinum that was centered at $a_0=3.916$ \AA.
We used $7$ reference geometries separated by $\Delta a = 0.07$ a.u. for all metals
except for platinum where we used $\Delta a = 0.0233$ a.u. for LDA and $\Delta a = 0.03$ a.u.
for PBEsol due to the presence of unstable strained configurations
with imaginary phonon frequencies at too large lattice constants.
We used 6 strained configurations for each type of strain with a strain interval $\delta \epsilon=0.005$. 
In total we computed the phonon dispersions on $126$ geometries for each material and functional in addition to
those computed on the reference configurations required to compute $a(T)$ and the thermal expansion. 
For the electronic calculation we computed the bands in all the reference configurations
(to include the effect on $a(T)$) and in all the strained ones (to include it in the TDECs). 
The \textbf{k}-point mesh for the electronic DOS calculations was $48 \times 48 \times 48$.
We use a Gaussian smearing with a smearing parameter of $0.01$ Ry. 
The inclusion of the electronic finite temperature effect in the thermodynamic quantities and in the 
TDECs is documented in the \texttt{thermo\_pw} user's guide.
In order to fit the free-energy as a function of the strain we used a polynomial of degree two because, although the use
of a fourth-degree polynomial could introduce some differences, it requires higher cutoffs to converge
the $T=0$ K ECs. To interpolate the ECs computed at the different reference configurations and calculate them at the 
temperature dependent geometry we use a fourth-degree polynomial. More informations about the
convergence tests are reported in the supplementary material.

The DFPT finite-temperature approach was applied to evaluate the ECs of platinum at the temperatures $T_{FD}=800$ K 
(with a Fermi-Dirac smearing $\sigma_{FD} \approx 0.005$ Ry) and $T_{FD}=1000$ K ($\sigma_{FD} \approx 0.0063$ Ry).
For this purpose the TDECs calculation was set in a single reference geometry with the lattice constant
at the considered temperature: 7.429 a.u. for $800$ K and 7.446 a.u. for $1000$ K.

\begin{figure}
\centering
\includegraphics[width=\linewidth]{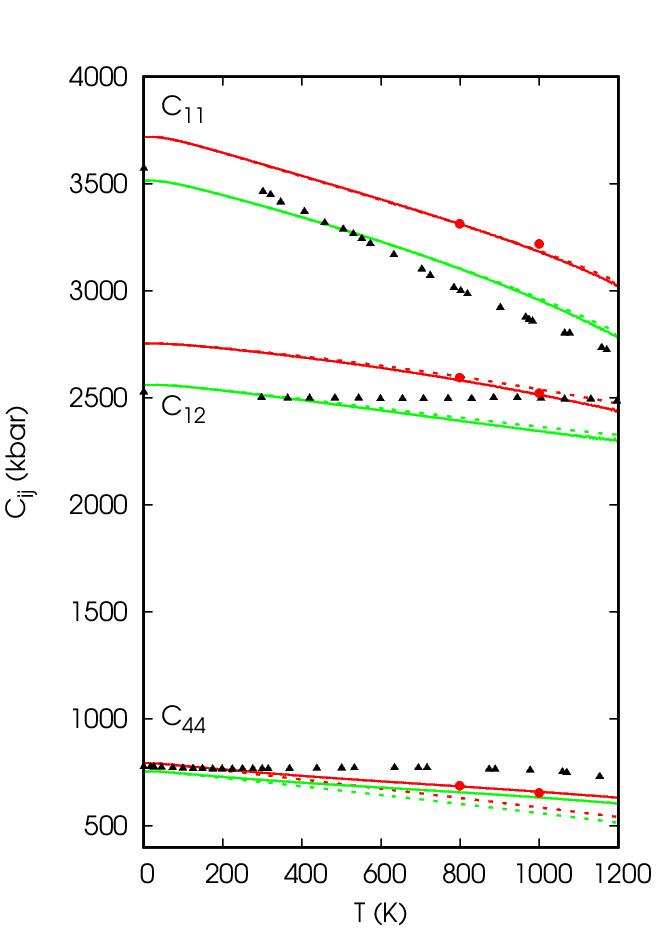}
\caption{Quasi-harmonic adiabatic elastic constants of platinum. LDA (red curves) is compared
         with PBEsol (green curves). The results obtained with the total free-energy (continuous line)
         are compared with those in which the contribution of electronic excitations is neglected (dashed lines).
         The red circles at $800$K and $1000$K are computed within the LDA by using the Fermi-Dirac occupations.  
         Experimental data are taken from Collard and McLellan~\cite{Collard_1992} (black triangles).}
\label{fig:pt}
\end{figure}

\section{Applications}
In Table 1, we report the zero temperature lattice constants, 
elastic constants, and bulk moduli of the four metals calculated with 
different exchange and correlation functionals. 
We compare them with experiments and previous theoretical works.
As usually found, PBE and PBEsol give slightly larger lattice parameters than LDA 
and smaller $T=0$ K ECs. PBEsol reproduces
the experimental ECs of palladium and platinum better than LDA and PBE
with errors smaller than $\approx 5 \%$ for all the ECs, compared 
to LDA errors in the range $10-18 \%$ in palladium and from $\approx 9 \%$ 
($C_{11}$ and $C_{44}$) to $\approx 12 \%$ ($C_{12}$) in platinum.
PBE has differences till to $\approx 15-16 \%$ for both materials.  

On the other hand, the experimental ECs of copper are well reproduced by PBE  
with errors equal or smaller than $3 \%$, compared with errors larger
than $15 \%$ for the other functionals. The LDA and PBEsol have almost 
the same accuracy in reproducing the $T=0$ K $C_{11}$ and $C_{44}$ of 
gold (with errors of $\approx 5 \%$ and $\approx 18 \%$, respectively), while
for $C_{12}$ PBEsol has an error of $\approx 3 \%$ and LDA an error 
of $\approx 10 \%$, PBE has errors of $\approx 22 \%$ in $C_{11}$ and 
$C_{12}$ and $\approx 40 \%$ in $C_{44}$.

The differences found for the LDA ECs of palladium with respect to those of Ref.~\cite{Liu_2011}
are $\approx 2 \%$ for $C_{11}$, $\approx 29 \%$ for $C_{12}$ and $\approx 10 \%$ for $C_{44}$.
For platinum, comparing our LDA ECs with the calculations of Ref.~\cite{Isaev_2008} we found differences
of $\approx 4 \%$ for $C_{11}$, $\approx 3 \%$ for $C_{12}$ and $\approx 8 \%$ for $C_{44}$;
while the comparison of our PBEsol ECs with the corrisponding PBEsol 
all-electrons calculation of Ref.~\cite{Jamal_2014} gives differences of $\approx 1 \%$ for $C_{11}$, 
$\approx 5 \%$ for $C_{12}$ and $C_{44}$. 
The differences found for the LDA-ECs of copper with
respect to those of Ref.~\cite{Wang_2009} are $\approx 28 \%$ for all ECs, while
the differences between our PBE ECs with the other GGA ECs of Refs.~\cite{Liu_2012, Hafner_2009} 
are smaller than $\approx 3 \%$ for $C_{11}$ and $C_{12}$ and $\approx 5-6 \%$ for $C_{44}$. 
Our PBEsol ECs of gold are close to the GGA estimates of Ref.~\cite{Wang_2009}
with differences of $\approx 5 \%$ for $C_{11}$ and $C_{12}$ and $\approx 3 \%$ for $C_{44}$.
For reference, in palladium, platinum, and gold we report GGA ECs found in literature. 
As expected, they are smaller than our LDA and PBEsol ECs. They shows 
some differences with our PBE estimates, which are, however, very close to the PBE ECs
reported in the Materials Project database and computed with the PAW method.

\begin{figure}
\centering
\includegraphics[width=\linewidth]{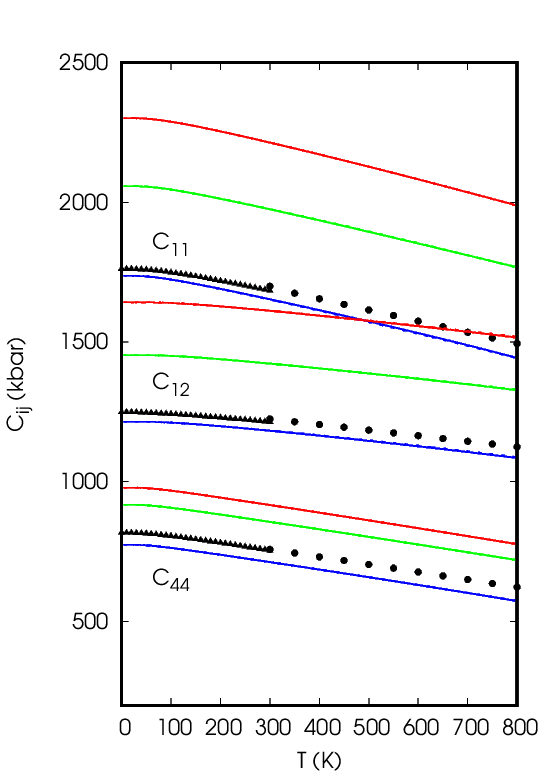}
\caption{Quasi-harmonic adiabatic elastic constants of copper. LDA (red curves), PBEsol (green curves) and PBE (blue) are
         compared. The results obtained with the total free-energy (continuous line) overlaps those in which
         the contribution of electronic excitation is neglected (dashed lines). Experimental data are taken from Overton and 
         Gaffney~\cite{Overton_1955} (black triangles) and Chang and Himmel~\cite{Chang_1966} (black circles).}
\label{fig:cu}
\end{figure}

\begin{figure}
\centering
\includegraphics[width=\linewidth]{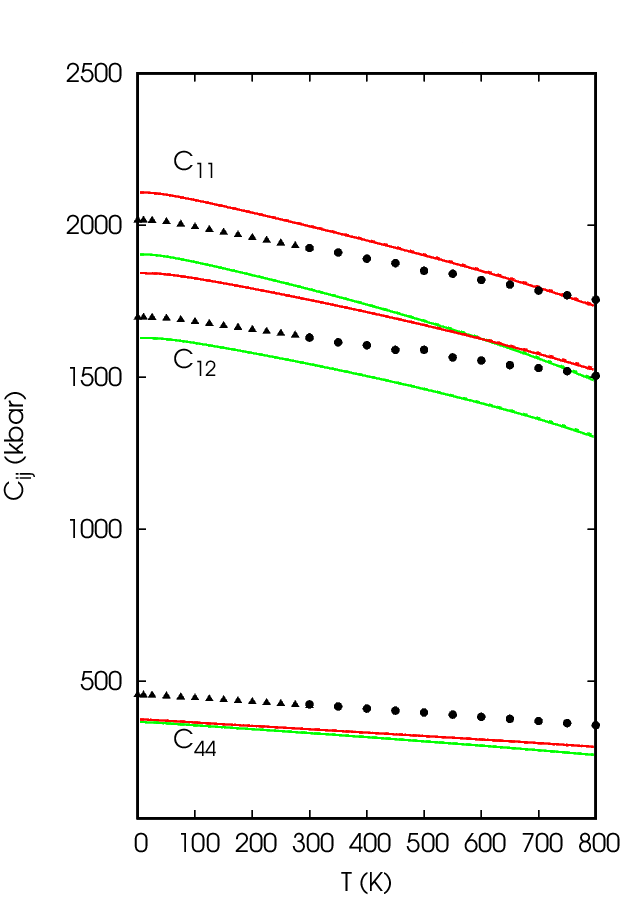}
\caption{Quasi-harmonic adiabatic elastic constants of gold. LDA (red curves) and PBEsol (green curves) are
         compared. The results obtained with the total free-energy (continuous lines) overlaps with those in 
         which the contribution of electronic excitation in neglected (dashed lines). Experimental data are 
         taken from Neighbours and Alers ~\cite{Neighbours_1958} (black triangles) and Chang and Himmel
         ~\cite{Chang_1966} (black circles).}
\label{fig:au}
\end{figure}

In Figs.~\ref{fig:pd}, \ref{fig:pt}, \ref{fig:cu}, \ref{fig:au} we report the TDECs of the 
four metals. For each plot we follow the same color-line convention: 
Red indicates LDA estimates, green PBEsol and blue, present only in copper, PBE. 
The continuous lines are computed by considering all contributions in the 
Helmholtz free-energy, while the dashed lines have been obtained neglecting
the contribution of the electron thermal excitations. Black points are 
the experimental
data. 

The TDECs of palladium are shown in Fig.~\ref{fig:pd}. Two experimental data-set
are reported. The variation of the experimental ECs of Ref. \cite{Rayne_1960} in the range $0-300$ K is $3 \%$ for
$C_{11}$, about $0 \%$ for $C_{12}$ and $-1 \%$ for $C_{44}$ with respect to the $0$ K ECs.
The corresponding LDA (PBEsol) variations in the same range of temperature are
$3.7 \%$ ($4.5 \%$) for $C_{11}$, $2 \%$ ($2 \%$) for $C_{12}$ and $3.8 \%$ ($3.8 \%$) for $C_{44}$.
In the range $T=300-1200$ K the variations in the data of Ref. \cite{Yoshihara_1987} are $7.3 \%$ for $C_{11}$, $6.4 \%$ for $C_{12}$ and $7.8 \%$ for $C_{44}$, while the corresponding
theoretical softenings are $15 \%$ for $C_{11}$, $13 \%$ for $C_{12}$, $11 \%$
for $C_{44}$ for both functionals.
Hence, both functionals give a similar temperature dependence. 
At room temperature the details of the experimental
anomalies are not reproduced and above room temperature the theoretical softening 
is larger than the experimental one. 
The electronic effect is very small in $C_{11}$ and $C_{12}$ but leads to 
appreciable modifications of $C_{44}$ which becomes closer to the experiment.  
As expected, the electronic contribution increases with temperature and
influences the softening of $C_{44}$ that without the electronic
effect would be $20 \%$ in the range $300-1200$ K. 

The TDECs of platinum are reported in Fig.~\ref{fig:pt}. From $T=0$ K
to $T=1200$ K the experimental ECs decrease of about $24 \%$ for $C_{11}$,
about $2 \%$ for $C_{12}$ and about $6 \%$ for $C_{44}$. The corresponding
theoretical softening for LDA (PBEsol) are $18 \%$ ($20 \%$) for $C_{11}$,
$11 \%$ ($10 \%$) for $C_{12}$ and $19 \%$ ($19 \%$) for $C_{44}$.
Hence, the variation of $C_{11}$ is well reproduced while the other 
two ECs' softenings are overestimated. 
As found for palladium, the electronic contribution is important and
improves the comparison with the experiment only for
the $C_{44}$ EC: without the electronic effect the softening
would be $\approx 30 \%$. 
However, it can not describe the anomalies.
In order to further investigate the trend of the ECs of platinum we also 
considered the effect of the electronic excitation on the phonon 
frequencies for the temperatures $T=800$ K and $T=1000$ K  
with the method explained in the previous section (results are reported in red
circles in the plot, only for the LDA case). 
The red points are exactly over our curves (apart for a slight deviation in the $C_{11}$ at $1000$ K).
This fact points out that the electron-phonon interactions that modify the phonon 
frequencies has negligible consequences on the TDECs.
We observe that our estimate of the electronic contribution is smaller 
than the one of Ref.~\cite{Keuter_2019} even for the $C_{44}$. In our calculation the largest 
contribution is the vibrational one which always decreases the ECs as the temperature 
increases. In Ref.~\cite{Keuter_2019} the vibrational contribution is described 
within a quasi-static framework which leads to an underestimation of the ECs' softening. 
Our results are qualitatively closer to those of Ref.~\cite{Liu_2011}
that finds a conventional temperature dependence for the $C_{44}$.

The results of copper are shown in Fig.~\ref{fig:cu}. Since the two different
experimental set of data are smoothly connected we report the whole softening
in the range of temperature $T=0-800$ K: $15 \%$ for $C_{11}$, 
$10 \%$ for $C_{12}$
and $24 \%$ for $C_{44}$. The corresponding theoretical softening for LDA (PBEsol,
PBE) are $13.5 \%$ ($14 \%$, $17 \%$) for $C_{11}$, $7.6 \%$ ($8.5 \%$, $10.5 \%$)
for $C_{12}$ and $21 \%$ ($21 \%$, $26 \%$) for $C_{44}$.
The temperature dependence is almost the same for LDA and PBEsol, slightly larger 
for PBE. The estimated softenings agree very well with the experiment. 
PBE results are the closest to the experiment, reflecting the trend of the 
$T=0$ K ECs shown in Table 1.  The electronic thermal excitations do not give any observable 
contribution in the whole range of temperature.

The TDECs of gold are shown in Fig.~\ref{fig:au}. As for copper we consider the
two experiments together: In the range of temperature from $T=0$ K to 
$T=800$ K the softenings are $13 \%$ for $C_{11}$, $11 \%$ for $C_{12}$
and $22 \%$ for $C_{44}$. The corresponding softening for LDA (PBEsol) are $18 \%$ ($22 \%$) 
for $C_{11}$, $17 \%$ ($20 \%$) for $C_{12}$ and $24 \%$ ($29 \%$) for $C_{44}$. 
The PBEsol softening is slightly larger than the LDA one and both functionals overestimate 
the experimental softening in particular for $C_{11}$ and $C_{12}$ with differences that increase from
room temperature onward. As in copper, the electronic thermal excitations 
do not give any observable contribution.

\section{Conclusions}

In this paper we investigated the TDECs of palladium,
platinum, copper, and gold by means of the QHA.
LDA and GGA give almost the same softenings but
differ in the zero temperature values. This work shows
that PBEsol is the best choice for palladium, platinum,
and gold. PBE is the best for copper.
Moreover, we addressed the role of the thermal electronic excitations 
and found a negligible effect on the TDECs of copper and gold,
and an improvement in the agreement with the experimental $C_{44}$
of palladium and platinum. Even though the computed softenings of all
ECs is in reasonable agreement with the experimental ones, we could not
reproduce the precise behavior of the anomalies in platinum and palladium.
In platinum, we further investigated the effects 
of the change of the phonon frequencies induced by the finite temperature
thermal electronic excitations, finding that it is small, even at the highest
studied temperature $T=1000$ K.
At the current stage the discrepancy between theory and 
experiment in palladium and platinum is not solved and might require either a 
more sophisticated functional than LDA or GGA or a method that includes the
anharmonic effects beyond the QHA approximation~\cite{Shulumba_2015}.
Since the experimental data are quite old, also a novel measurement of the TDECs 
could be useful.

\ack

Computational facilities have been provided by SISSA through its Linux
Cluster and ITCS and by CINECA through the SISSA-CINECA 2019-2021
Agreement.

\section*{References}

\bibliography{unsrt}{}
\bibliographystyle{unsrt}

\clearpage

\end{document}


\title[]{Supplementary material for Quasi-harmonic thermoelasticity of 
palladium, platinum, copper, and gold from first principles.}




%

\begin{table*}[b] \centering
\caption{Convergence test of zero-temperature ECs vs. energy cutoffs with unconverged
smearing parameter ($\sigma=0.02$ Ry) and {\bf k}-points grid ($16 \times 16 \times 16$) computed within LDA. The ECs and the bulk moduli 
$B$ are in kbar. The bulk modulus is $B=\frac{1}{3}(C_{11}+2C_{12})$.
}
\begin{tabular}{lccccc}
\hline
\multicolumn{1}{c}{ecut\_wf (ecut\_rho)} \ \ & $a_0$ \ \ & \multicolumn{1}{c}{$C_{11}$} \ \ & \multicolumn{1}{c}{$C_{12}$} \ \ & \multicolumn{1}{c}{$C_{44}$} \ \ & \multicolumn{1}{c}{$B$} \\
\hline
\hline
Palladium \\
\hline
45(300) & 7.275 & 2725 & 2071 & 874 & 2289 \\
{\bf 60(400)} & 7.274 & 2699 & 2062 & 864 & 2274 \\
100(800) & 7.274 & 2723 & 2069 & 868 & 2287 \\
\hline
\hline
Platinum \\
\hline
{\bf 45(300)} & 7.373 & 3819 & 2740 & 891 & 3100 \\
60(400) & 7.373 & 3811 & 2748 & 892 & 3102 \\
100(800) & 7.372 & 3818 & 2743 & 892 & 3101 \\
\hline
\hline
Copper \\
\hline
45(1200) & 6.651 & 2359 & 1651 & 1005 & 1887 \\
{\bf 60(1200)} & 6.651 & 2376 & 1656 & 1008 & 1896 \\
60(1600) & 6.651 & 2373 & 1660 & 1010 & 1898 \\
100(2000) & 6.651 & 2372 & 1668 & 1010 & 1903 \\
\hline
\hline
Gold \\
\hline
45(300) & 7.663 & 2162 & 1846 & 428 & 1951 \\
{\bf 60(400)} & 7.662 & 2138 & 1860 & 426 & 1953 \\
100(800) & 7.662 & 2168 & 1842 & 427 & 1951 \\
\hline
\end{tabular}
\label{table2}
\end{table*}

\section{Convergence tests}
In this section we discuss some convergence tests of the zero-temperature elastic constants (ECs) with 
respect to the energy cutoffs, smearing parameters and {\bf k}-points meshes.
The ECs are computed by fitting the energy-strain curves. We used 6 value of the strain: a reasonably 
compromise between accuracy and the need to reduce the number of phonon calculations. 
To fit them, a second-degree polynomial was selected that, in the previous literature, was found to be
the best polynomial to give the ECs for a small set of points (Computer Physics Communications {\bf 184}, 1861
(2013) and Physical Review B {\bf 85} 064101 (2012)).
A higher polynomial degree, like fourth, requires larger energy cutoffs to converge the ECs.
We discuss below the choice of the polynomial and the interval $\Delta \epsilon$
between strained configurations. For the following tests we used $\Delta \epsilon = 0.005$
which was found to be a good choice for many materials (see, for example, Refs. [16-18] of the paper).

In Table~\ref{table2} the ECs as well as the lattice parameters and the bulk moduli are reported for increasing 
cutoffs. We started with the cutoffs needed to converge the phonon frequencies (Ref. [8] of the paper). 
We ensure that $C_{11}$ and $C_{12}$ are converged within at least $\approx 20$ kbar
and a few kbar for the $C_{44}$. The chosen cutoffs are written in bold type. The tests in the tables are
for the LDA functional but the final cutoffs were chosen also considering the tests with the other
functionals.

\begin{figure}
\centering
\includegraphics[width=1.2\linewidth]{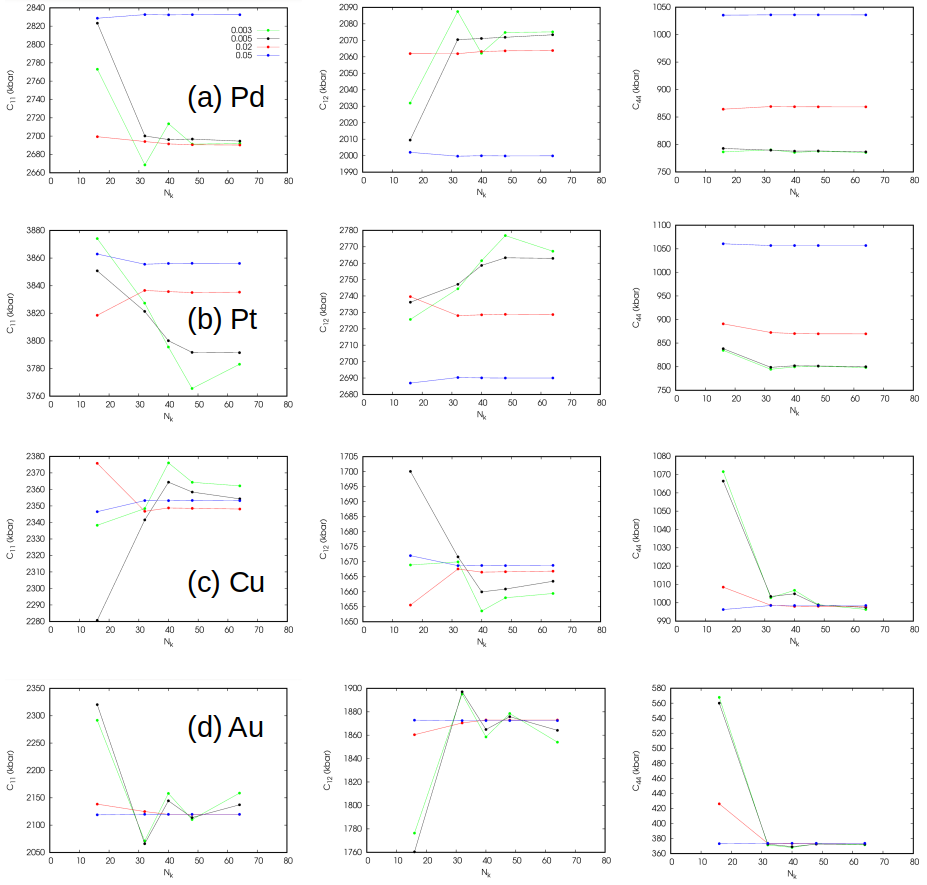}
\caption{Tests of the zero temperature ECs vs. the smearing parameters
and the k-points grid $N_k \times N_k \times N_k$ for each material. The color indicates the smearing
parameter: $\sigma=0.003$ Ry (green), $\sigma=0.005$ Ry (black), $\sigma=0.02$ Ry (red) and $\sigma=0.05$ Ry (blue)
The final set of parameters is $\sigma=0.005$ Ry for palladium and platinum, $\sigma=0.02$ Ry for copper and gold 
and $N_k=40$ for all systems.}
\label{kpoints}
\end{figure}

With the converged cutoffs we test the smearing parameter and the {\bf k}-points grids
for each material. These tests are reported in Fig.~\ref{kpoints}. The final set of
ECs are those obtained with the parameters declared in the paper ($\sigma=0.005$ Ry for 
palladium and platinum, $\sigma=0.02$ Ry for copper and gold and $N_k=40$ for all systems).

In order to further check the ECs we compute them as the derivatives of the stress with respect
to the strain. For this purpose we compute directly the stress (by using the stress theorem
as implemented in $QuantumESPRESSO$ and explained in Physical Review B {\bf 32} 3780 (1985)) in four strained configurations around 
the equilibrium and then fit with respect to the strain with a second-degree polynomial and
evaluate the analytic first derivative. The results are collected in Table~\ref{table3}.
The ECs computed within the two methods are in good agreement with differences usually less than $1 \%$,
only in the $C_{11}$ and $C_{12}$ ECs of copper the differences are $\approx 2 \%$ but the agreement
is still satisfactory. 
\begin{table*}[b] \centering
\caption{Comparison between ECs computed from the second derivatives of the total energy and those
computed from the first derivatives of the stress with respect to the strain. 
We used the converged parameters discussed in Table~\ref{table2} and Fig.~\ref{kpoints}.
The functional is LDA. The ECs and the bulk moduli $B$ are in kbar.
The bulk modulus is $B=\frac{1}{3}(C_{11}+2C_{12})$.
}
\begin{tabular}{lcccc}
\hline
\multicolumn{1}{c}{Method} \ \ & \multicolumn{1}{c}{$C_{11}$} \ \ & \multicolumn{1}{c}{$C_{12}$} \ \ & \multicolumn{1}{c}{$C_{44}$} \ \ & \multicolumn{1}{c}{$B$} \\
\hline
\hline
Palladium \\
\hline
Energy & 2696 & 2071 & 788 & 2279 \\
Stress & 2705 & 2070 & 782 & 2282 \\
\hline
\hline
Platinum \\
\hline
Energy & 3800 & 2759 & 802 & 3106 \\
Stress & 3788 & 2754 & 790 & 3099 \\
\hline
\hline
Copper \\
\hline
Energy & 2349 & 1666 & 998 & 1894 \\
Stress & 2293 & 1625 & 998 & 1848 \\
\hline
\hline
Gold \\
\hline
Energy & 2120 & 1873 & 373 & 1955 \\
Stress & 2130 & 1849 & 374 & 1943 \\
\hline
\end{tabular}
\label{table3}
\end{table*}


\clearpage

\begin{figure}[b]
\centering
\includegraphics[width=1.2\linewidth]{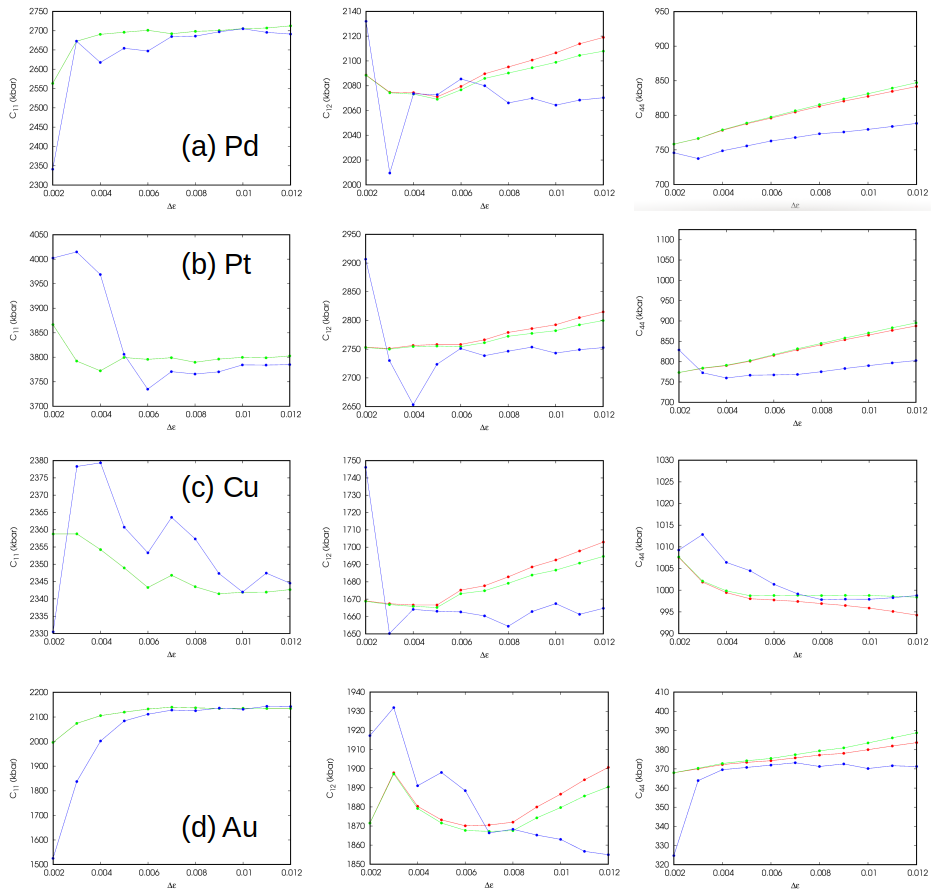}
\caption{Tests as a function of the interval $\Delta \epsilon$ between two strained configuration for each
EC of each material. The number of points is fixed at 6. The maximum applied strain can be obtained as $\epsilon_{max}=2.5 \Delta \epsilon$. Second degree (red), third degree (green) and fourth degree (blue) fitting polynomials are compared. The value chosen for the calculation is $\Delta \epsilon=0.005$.}
\label{poly}
\end{figure}

Finally, we check the parameters $\Delta \epsilon$ and the degree of the polynomial in Figure~\ref{poly}.
The second degree polynomial (red) provides ECs equal or very similar ECs to those computed with the third 
degree polynomial (green): their values are the same in a reasonably large interval of $\Delta \epsilon$
around $\Delta \epsilon=0.005$. While the forth degree polynomial (blue) reaches the plateau for larger values 
of $\Delta \epsilon$. 
In any case the differences are usually very small, the largest differences between
the second and fourth degree polynomial are found in the $C_{44}$ ECs of palladium and platinum that are
however of the order of $32-34$ kbar with $\Delta \epsilon = 0.005$. This difference reduces if the ECs obtained
from the fourth degree polynomial are computed with a larger cutoff. For example in palladium becomes $19$ kbar
with a cutoff equal to $110$ Ry and $800$ Ry for the wave functions and charge density, respectively.